TECHNICAL REPORT | MARCH 2026

# Global Cybercrime Damages

## A Baseline for Frontier AI Risk Assessment

Kamile Lukosiute, John Halstead, Luca Righetti

GovAI

## Executive Summary

AI companies and governments have raised concerns about frontier AI systems enabling cyber-attacks and cybercrime. Some have expressed interest in defining so-called capability thresholds: If there is evidence of AI models creating severe harm via cybercrime, then concrete mitigations should be triggered. But it remains unclear where to draw the line. Should it be when AI is believed to increase cybercrime by 10%, or double it, or something else? To make that judgment, we first need to know the scale of cybercrime today.

A fundamental challenge is that current estimates of global cybercrime damages vary widely, from tens of billions to tens of trillions of dollars, with little systematic evaluation of their reliability. The existing literature notes that many headline estimates from cybersecurity vendors may be inflated due to commercial incentives, while others rely on questionable methodologies. This uncertainty makes it difficult for AI developers to assess at what point future capabilities might breach their damage thresholds – or for policymakers to allocate resources effectively.

This report establishes a more rigorous baseline by surveying 27 different estimates and critically evaluating their methodologies. We adopt a taxonomy of cybercrime that distinguishes cyber-dependent crimes (e.g. hacking of computers) from cyber-assisted crimes (e.g. internet-enabled fraud). Our focus is on quantifiable economic damages:

- **Direct losses:** money stolen or extorted
- **Response costs:** incident remediation and investigation
- **Defense spending:** prevention measures

We exclude harder-to-measure indirect costs such as intellectual property losses and reputational damage, and additional social harms such as national security considerations.

We construct a composite estimate from three independent sources of evidence: (1) a nationally representative business victimisation survey from the UK government, scaled globally; (2) individual victimisation data from a US academic survey, scaled globally; and (3) global cybersecurity spending figures as a proxy for defense costs. Large-sample victimisation surveys capture losses directly from victims, which avoids both the reporting bias common in estimates and the heavy modeling assumptions required by approaches relying on macroeconomic estimates.

**We find:**

- Individual direct losses: ~**$200 billion** annually
- Business direct and response costs: ~**$200 billion** annually
- Additional defense spending: **~$100 billion** annually
- Total: ~**$500 billion** annually
    - 90% confidence interval: **$100 billion to $1 trillion**

**Implications for AI Risk Management**

**If this report's estimates are correct, a 20% increase would cross commonly cited damage thresholds.** If current cybercrime damages approach $500 billion annually, an AI-driven increase of ~20% could add $100 billion or more, which would reach thresholds some companies identify as warranting additional risk mitigations. This is a relatively modest increase that could occur through efficiency gains in existing attack methods without requiring qualitatively new AI capabilities. Lower thresholds require correspondingly smaller uplifts: reaching $10 billion in additional damages would require only ~2% uplift, and reaching $1 billion would require only ~0.2%.

**However, data on cybercrime damage is too incomplete and ambiguous for incremental AI-driven increases to be directly detectable.** There is no single data source that accurately captures total damages due to pervasive under-reporting, evolving crime definitions, and measurement inconsistencies. This means that even after-the-fact estimates of models' contributions to cybercrime damage will necessarily be uncertain and require the use of multiple forms of evidence.

**Defensive adaptation also complicates net impact assessments.** AI can substantially enhance defensive capabilities, not just offensive ones. The net effect of AI could be positive on the cybercrime landscape, but future modeling and data collection is needed to confirm this.

**Companies may also have regulatory obligations to consider potential cybercrime damage enabled by their AI systems.** Companies' safety frameworks typically focus on discrete events (e.g. individual cyberattacks) that cause large-scale harm. Cybercrime causes *cumulative* harm through many distributed incidents rather than a single catastrophic event. Yet regulatory obligations may not require this instantaneous framing. The EU AI Act's Code of Practice identifies "enabling large-scale sophisticated cyber-attacks" as a systemic risk requiring assessment and mitigation, which could encompass AI-enabled scaling of many smaller attacks, not just single catastrophic events.

> **Areas for Future Research**
>
> - **Developing indicators beyond aggregate cost estimates**, such as monitoring for AI-generated code in malware samples, tracking criminal forum discussions of AI tool adoption, and observing labor market changes in fraud operations
>
> - **Understanding the net impact of AI defenses** across potential victims, including under-resourced government agencies, critical infrastructure providers, and organizations in developing countries
>
> - **Developing harmonized data** across multiple countries to reduce reliance on GDP scaling from Western nations
>
> Ultimately, although limitations remain, this report narrows plausible estimates of cybercrime damage from spans of multiple orders of magnitude to a more confident baseline. Our systematic methodology provides a foundation for more informed decision-making about AI-related cybercrime risks.

# Global Cybercrime Damages
## A Baseline for Frontline AI Risk Assessment

Kamile Lukosiute, John Halstead, Luca Righetti

## Introduction

### Background and Motivation

AI companies, governments and other experts have raised concern about the potential misuse of frontier AI systems for cyber-attacks,[1] for example, by "scaling the volume of existing cyberattacks."[2] The EU AI Act's Code of Practice for general-purpose AI likewise identifies "enabling large-scale sophisticated cyber-attacks" as a systemic risk and identifies the need to perform risk modeling for this particular threat.[3] Law enforcement organizations have also expressed concern about AI's potential to enable not only cyber-attacks by improving "criminal methods and scripts"[4] but also to enable sophisticated social engineering and cyber fraud.[5] Given this spectrum of societal concerns, this paper estimates economic damages from cybercrime broadly, which includes both cyber-attacks and other cyber-enabled fraud, providing a baseline against which to assess AI-enabled threats.

Despite attention from law enforcement, policymakers, and AI developers, the risk model of AI-enabled cybercrime remains poorly understood. Some Frontier AI Safety policies focus on single catastrophic events, but the nature of cybercrime is cumulative: many small, common events rather than one dramatic incident. Even taking conservative readings of existing estimates, yearly cybercrime costs (in the hundreds of billions) are 10x the contribution of any single incident – the estimated cost of WannaCry, for example, is on the high end $10B.[6] This cumulative burden could pose a systemic risk: an increase in the total volume of cyber-attacks and cyber fraud disrupts the present offense-defense balance, strains defensive resources, and erodes trust in digital systems.

Threat modeling of AI-enabled cybercrime is therefore warranted. One approach is to estimate expected damages as a function of two inputs: the AI-enabled percentage increase in cybercrime, and the baseline level of current damages (Figure 1).

---

[1] OpenAI, "Preparedness Framework," last updated April 15, 2025; Meta, "Frontier Ai Framework," n.d.; Anthropic, "Responsible Scaling Policy," effective March 31, 2025; Google, "Frontier Safety Framework Version 3.0," September 22, 2025; International AI Safety Report 2026, February 3, 2026, 72–78.
[2] OpenAI, "Preparedness Framework."
[3] European Union, "EU AI Act: General-Purpose AI Code of Practice," 2025, Appendix 1.4.
[4] Europol, "IOCTA, Internet Organized Crime Threat Assessment 2024," *Europol*, 2024, 33.
[5] Europol, "IOCTA."
[6] Tom Johansmeyer, "Surprising Stats: The Worst Economic Losses from Cyber Catastrophes," The Loop, March 12, 2024.

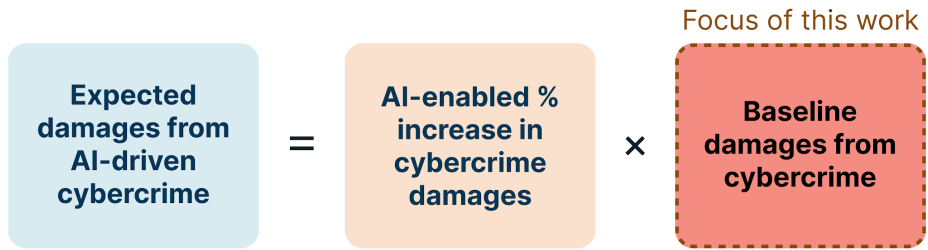

**Figure 1 | Estimating Damages from AI-driven Cybercrime.** The expected increase in damages from AI-enabled cybercrime can be estimated by estimating the baseline damages from cybercrime and the % increase in cybercrime damages from AI.

**This paper produces a baseline estimate of global cybercrime damages to inform threat modeling of AI-enabled cybercrime.** A reliable estimate can help AI companies determine whether and how to define cyber capability thresholds in their Frontier AI Safety Policies, which we elaborate on in the Discussion. We leave estimation of the AI-enabled uplift to future empirical work on how frontier models affect attacker and defender capabilities.

Producing an accurate estimate of cybercrime damages is challenging. Definitions of cybercrime vary, data quality is poor, and victims underreport incidents.[7] Moreover, many prominent estimates by cybersecurity vendors may be inflated due to vendors' incentives to gain publicity through large damage figures.[8] By surveying and addressing these methodological challenges, we aim to guide decision makers through this literature and produce a more robust estimate given the inherent uncertainty. We note that even arriving at a more reliable order of magnitude – whether cybercrime costs ~$10 billion, ~$100 billion, or ~$1 trillion – can be highly informative for decision makers.

---

[7] U.S. Department of Justice, "Victimizations Not Reported to the Police, 2006–2010" (*Bureau of Justice Statistics*, August 2012); The Council of Economic Advisers, "The Cost of Malicious Cyber Activity to the U.S. Economy" (Executive Office of the President of the United States, February 2018).
[8] Bruce Sterling, "Global Cybercrime. Costs a Trillion Dollars. Maybe 3," *WIRED*, July 19, 2017; R. Anderson et al., "Measuring the Changing Cost of Cybercrime," The 18th Workshop on the Economics of Information Security (WEIS), January 2019; Cybersecurity and Infrastructure Security Agency, "Cost of a Cyber Incident: Systematic Review and Cross-Validation," October 2020; John Leyden, "Why Is Mi2g so Unpopular?," *The Register*, November 21, 2002; Air Force Research Library, "Economic Analysis of Cyber Security," July 2006, 25.

## Report Outline

This report proceeds as follows:

- In [Section 2](), we define cybercrime and cybercrime damages, establishing a taxonomy that distinguishes cyber-dependent crimes (attacks on systems) from cyber-assisted crimes (technology-enabled fraud), and categorising costs into direct losses, response costs, defense spending, and indirect impacts.

- [Section 3]() reviews existing methods for estimating global cybercrime damages, examining 27 sources that employ approaches ranging from victimisation surveys to cryptocurrency flow analysis, and critically evaluates two commonly cited estimates in more depth: FBI complaint data and economic models by RAND.

- [Section 4]() presents our own composite estimate of global cybercrime damages, grounded in business victimisation survey data from the UK, individual victimisation data from the US, and cybersecurity spending estimates to arrive at a total range of $200 billion to $550 billion annually.

- [Section 5]() discusses the implications of our findings for AI safety policies, emphasizing that measuring AI's impact on cybercrime will require building alternative measures of uplift because the error bounds in aggregate cost estimates are too large to see an effect.

- The [appendices]() provide detailed calculations, ransomware cost analysis, and additional estimates of scam center criminal revenues.

## Defining Cybercrime and Cybercrime Damages

To measure cybercrime damages, we first need:

- **A definition of cybercrime:** What exactly does a crime need to involve to be classed as 'cybercrime'?

- **A definition of cybercrime damages:** How socially damaging is a given amount of cybercrime to society?

### Defining Cybercrime

There is no commonly agreed upon definition of cybercrime,[9] a problem that partly motivated Congress to pass the "Better Cybercrime Metrics Act."[10] As part of the congressional effort to improve cybercrime measurement, the National Academies of Science developed a taxonomy of cybercrime distinguishing cyber-dependent crimes (those targeted specifically against machines and data) from cyber-assisted crimes (other types of crime, such as fraud, enabled by technology). This taxonomy is shown in [Figure 2](#).

We adopt this framework for systematically categorising cybercrime throughout this work. This framework is focused on the method and target of the attacks to define what constitutes a cybercrime. Therefore, attacks that are ideological in motivation but ultimately lead to illegal and unauthorized access to networks would be included. In this way, damages from state-sponsored espionage as well as damages from hacktivism are included in our estimates.

---

[9] United States Government Accountability Office, "Cybercrime: Reporting Mechanisms Vary, and Agencies Face Challenges in Developing Metrics," June 2023.
[10] Better Cybercrime Metrics Act, Public L. No. 117–116, 136 Stat. 1180 (2022).

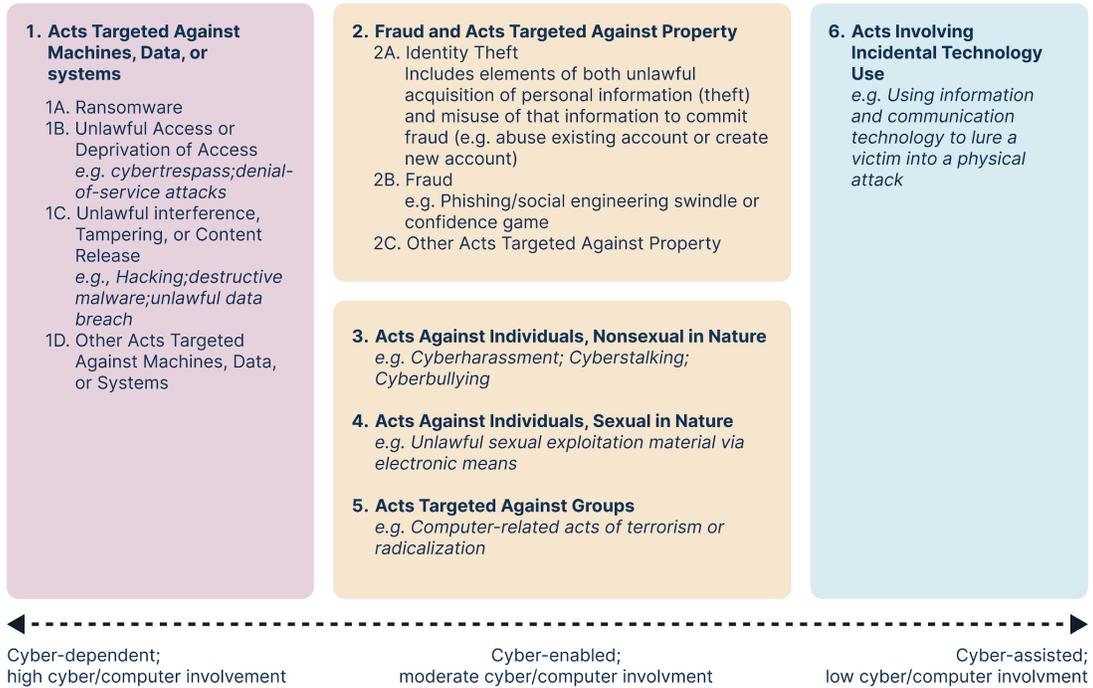

**Figure 5-1** Schematic of proposed classification of cybercrime.
SOURCE: Panel generated.

**Figure 2 | Cybercrime Classification and Measurement.** Reproduced from: "Cybercrime Classification and Measurement", NAS, 2025, pg. 9

This work focuses exclusively on categories (1) and (2), where established methods exist for estimating economic damages with reasonable precision. Categories (3) through (6) have real costs to society but their economic damages are more difficult to quantify reliably. Other work on cybercrime uses different and varying definitions, and it is important to bear in mind that different estimates of cybercrime damages may often in part reflect differences in the underlying definition of 'cybercrime' being used.

**Box 1. Notes on Scope Choices**

In discussions of AI within the AI community, "cybercrime" is often equated with, or primarily concerned with, cyber-attacks, particularly the danger of AI enabling technical exploits. For example, the International AI Safety Report discusses AI's increasing ability to "discover and patch exploitable software flaws,"[11] while the Appendix 1.4 of the EU AI Code of Practice identifies "risks from enabling large-scale sophisticated cyberattacks" as a key systemic risk.[12] This raises the question: why include category (2), which consists primarily of fraud, in this analysis?

---

[11] International AI Safety Report 2025, "First Key Update: Capabilities and Risk Implications," October 2025, 18.
[12] European Union, "EU AI Act: General-Purpose AI Code of Practice."

> There are two reasons. First, law enforcement organizations express great concern about AI-enabled fraud operations, and have already observed such cases.[13] Second, we include category (2) because fraud is typically how the harms from data breach attacks are ultimately realized by consumers. A substantial portion of fraud relies on information acquired through data breaches, and the majority of data breaches are conducted specifically to steal personally identifiable information for resale and ultimate use in fraud operations.[14] While not all cyber fraud requires breach-acquired data (such as "pig butchering" scams), category (2) losses in the NAS taxonomy should be at least partly attributed to cybercrime costs borne not by breached institutions, but by downstream individuals.

### Defining Cybercrime Damages

In this subsection, we explain how we will define cybercrime damages, i.e. how much financial and social damage a given amount of cybercrime causes. This is a conceptually challenging issue. For example, a ransomware attack on a company may cause the company to send an amount of money to criminals, but the damages of the attack are often larger. For example, the attack might also cause reduced production and thereby decrease revenues; it might cause the company and other companies to invest more in cybersecurity; it might reduce the stock price of the affected company or affect their reputation, and so on.

In Table 1, we set out our definitions of different types of cybercrime damages. These definitions draw on, but are in some respects different to, definitions given in several different sources.[15]

---

[13] Europol, "IOCTA"
[14] Kosinski, Matthew, "What is a data breach?" IBM, n.d.
[15] The FBI's IC3 report focuses primarily on direct financial losses – currency leaving a victim's account and going to a criminal. See FBI, "Internet Complaint Crime Center." The UK government's cyber security breaches survey expands this to include "short-term direct costs," "long-term direct costs," and "indirect costs," encompassing clean-up expenses and lost asset value. See DSIT, "Cyber Security Breaches Survey 2025," updated June 19, 2025. Anderson et al. include the costs of defense and "prevention efforts," including "security products such as spam filters and antivirus; security services provided to individuals, such as awareness raising; security services provided to industry," etc.

| Cost Type | Definition |
|---|---|
| Direct | • currency leaving the victim's bank account or crypto wallet and going to the account/wallet of a criminal (with an error of exchange and laundering costs). Includes money that people were scammed out of, was stolen, or paid willingly as ransom. |
| Response | • payments to external IT consultants or contractors to investigate or fix the problem<br>• the cost of new or upgraded software or systems<br>• recruitment costs if you had to hire someone new<br>• the cost of any devices or equipment that needed replacing<br>• any legal fees, insurance excess, fines, compensation, or PR costs related to the incident<br>• Includes money that people were scammed out of, was stolen, or paid willingly as ransom. |
| Defense | • incident prevention costs (personnel and equipment) |
| Indirect | • business interruptions (time when staff could not do their jobs, lost business revenue directly attributable to attack)<br>• inability of citizens to access public services (government sector equivalent of "lost revenue")<br>• death, or loss of life quality (healthcare sector equivalent of "lost revenue")<br>• lost revenue due to reputational costs<br>• lost revenue (incl. by other organizations) due to loss of trust in safety of online services<br>• value of lost IP<br>• Erosion of trust in digital systems |

**Table 1 | Definitions of Cost Categories.** This table defines four categories of economic damages resulting from cybercrime: direct costs (money transferred to criminals), response costs (post-incident remediation expenditures), defense costs (preventive security investments), and indirect costs (downstream economic impacts). We estimate direct, response, and defense costs; indirect costs are excluded.

For the purposes of this work, we only attempt to estimate global direct, response, and defense costs.

We exclude what we call 'indirect costs' because estimating these costs, including intellectual property losses and expected future revenue losses from reputational damage, is more challenging than estimating direct, response, and defense costs. Methods for estimating indirect costs are more likely to rely on subjective judgment.[16] In economic cyber espionage, for example, even when theft is discovered, victims face protracted uncertainty in determining what was stolen, how it might be used, and the resulting financial impact to their competitive position.[17]

Even those estimates that attempt to use more objective measures, such as abnormal returns[18] or studying litigated cases of corporate espionage,[19] remain fraught due to the statistical difficulties of attributing firm-level outcomes to specific events.[20] These losses, as well as other indirect costs such as reduced willingness to engage in online banking due to fraud concerns, represent real and important impacts. However, given the measurement challenges, we leave rigorous estimation of these costs for future work. We also exclude other non-financial costs, such as national security costs.

In the next section, we discuss the viable methods we've identified for measuring these categories of costs against individuals and businesses/institutions due to cybercrime.

---

[16] William Akoto and Trey Herr, "Primer on the Costs of Cyber Espionage," American University Center for International Service, October 3, 2024; The Council of Economic Advisers, "The Cost of Malicious Cyber Activity to the U.S. Economy," Figure 1.
[17] Tom Johansmeyer, "Insuring the Unseen: Closing the Protection Gap in Economic Cyber-Espionage" (American University CSINT Policy Paper Series, Fall 2025).
[18] For example, see The Council of Economic Advisers, "The Cost of Malicious Cyber Activity to the U.S. Economy."
[19] Dan Ciuriak and Maria Ptashkina, "Quantifying Trade Secret Theft: Policy Implications" (CIGI Papers No. 253, May 2021).
[20] S. P Khotari and Jerold B. Warner, "Econometrics of Event Studies," Working Paper (Center for Corporate Governance, Tuck School of Business at Dartmouth, May 19, 2006).

# Reviewing Methods for Estimating Global Cybercrime Damages

To identify and produce estimates of global cybercrime damages, we carried out a rapid review of the following sources:

- Academic literature
- Reports and data sources from national governments and intergovernmental organizations
- Think tank reports
- Reports and data sources from cybersecurity companies and insurers.

We also consulted external experts to ensure completeness.

Our summary [sheet](#) collects 27 different estimates of different subsets of cybercrime damages. 7 of these estimates are produced directly by the relevant source, while 20 are our own Fermi estimates inferred from different data sources and based on various uncertain assumptions. [Table 2](#) shows all the sources we considered, grouped by method type.

There are 6 shared common methodology types:

**Aggregate consumer reports.** These estimates are based on cumulative losses reported to national centralized government agencies that collect crime reports for cybercrime and fraud, which we scale to world GDP with uncertainty bounds created using estimates of under-reporting rates.

**Victimisation surveys.** To better calibrate under-reporting, researchers and national bodies ask random, large samples of businesses or individuals if they have been victims of a cybercrime; we can then scale to the full population using average loss and incidence rate estimates.

**Illicit cryptocurrency flows.** By tracking inflows into known crypto wallets of cybercriminals and assuming all inflows are due to cybercrime activity, it is possible to estimate direct losses in crypto payments.

**Scammer revenue reports.** By tracking the daily revenues of scammers via self-reported earnings, it is possible to estimate the global earnings of scammers, which provides an estimate of direct losses due to scamming.

**Economic modeling.** These estimates are based on extrapolations of individual hacking event data to the global economy.

**No published methods.** We identify several sources providing estimates of cybersecurity spending or other cybercrime costs that do not publish a methodology.

| Method Type | Methodology Issues | Primary Source | Individual or business cost estimate | Loss Types Assessed | Is the final estimate our own or reported directly in the original source? |
|---|---|---|---|---|---|
| Infer from victimisation | Incomplete taxonomy coverage, absence of verification. | [Norton (2023)](#) | Individual | Individual | Our own |
| | | [Breen et al. (2022)](#) | Individual | Individual | Our own |
| | | [AIC (2024)](#) | Individual | Individual | Our own |
| | | [Riek and Böhme (2018)](#) | Individual | Individual | Our own |
| | | [FTC Fraud Survey (2019)](#) | Individual | Individual | Our own |
| | | [NCVS Financial Fraud Survey (2021)](#) | Individual | Individual | Our own |
| | | [NCVS Identity Theft (2023)](#) | Individual | Individual | Our own |
| | | [European Commission (2022)](#) | Business | Business | Our own |
| | | [Accenture (2019)](#) | Business | Business | Our own |
| | | [DSIT (2025)](#) | Business | Business | Our own |
| | | CBS (2023) | Business | Business | Our own |
| Infer from Aggregated Consumer Reports | Incomplete taxonomy coverage, absence of verification, under-reporting bias | UNODC (2024) | Individual | Direct | Our own |
| | | FTC Consumer Sentinel Report (2025) | Individual | Direct | Our own |
| | | Report Fraud (2025) | Individual, business | Direct | Our own |
| | | IC3 (2025) | Individual, business | Direct | Our own |
| | | Meurs (2024) and Meurs et al. (2025) | Business | Direct, response | Our own |
| Infer from illicit cryptocurrency flows | Incomplete taxonomy coverage | Chainalysis (2025) | Individual, business | Direct | Our own |
| | | Griffin and Mei (2024) | Individual, business | Direct | Our own |
| Infer from economic modeling | Incomplete taxonomy coverage, severity selection bias | RAND (2018) (Model 1) | Business | Direct, response, indirect | Reported |
| | | RAND (2018) (Model 2) | Business | Direct, response, indirect | Reported |
| | | RAND (2018) (Model 3) | Business | Direct, response, indirect | Reported |
| | | CEA (2018) | Business | Direct, response, indirect | Our own |
| Infer from scammer | Incomplete taxonomy | Various reports from | Individual | Direct | Our own |
| revenues in human trafficking reports | coverage | trafficking victims | | | |
| **No published methods** | Incomplete taxonomy coverage | **eSentire (2024)** | Business | Direct, response, indirect | Reported |
| | | **Forrester (2025)** | Business | Response, defense | Reported |
| | | **Gartner (2025)** | Business | Response, defense | Reported |
| | | **IDC (2025)** | Business | Response, defense | Reported |

**Table 2 | Estimates of global cybercrime costs.** Summary table of sources and estimates of global cybercrime costs considered in this work

**Elaborating on Methodology Issues**

The methods we review and develop ourselves all face an overlapping set of issues. To conceptualise where these issues arise, consider a framework for how one would ideally measure total cybercrime damages.

We can express total economic damages as the sum, across all crime types, of three terms: the average cost per incident, the relevant victim population, and the victimisation rate (Figure 3).

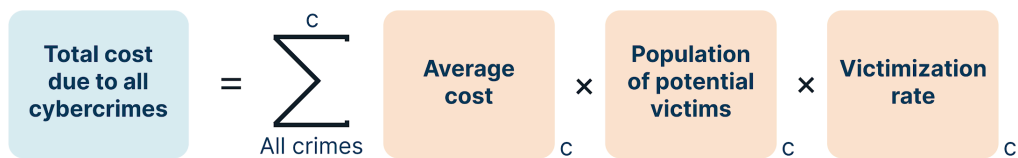

**Figure 3 | Cybercrime Cost Framework.** A conceptual framework for estimating the total cost of all cybercrimes. We use this framework to discuss the limitations of methods commonly used to derive global estimates.

Inaccurate estimation of any term in this equation leads to unreliable estimates.

First, many methods cover only some crimes within the summation and are unclear about which are included; we call this problem incomplete coverage. For example cryptocurrency analysis provides payment data,[21] but without knowing what fraction of total cybercrime payments flow through cryptocurrency, we cannot assess total cybercrime damages.

Second, estimates of the average cost may be biased in two ways. Lack of verification can inflate reported costs: victims typically report losses at the time of the incident, yet may subsequently recover a portion through chargebacks, insurance, or law enforcement intervention. Alternatively, when surveys do not clearly define "cost," victims may interpret it inconsistently. Without independent verification of victim reports, it is challenging to assess the accuracy of resulting estimates. In addition, average cost estimates derived from event databases suffer from severity selection bias: these databases capture only highly publicized or mandatorily disclosed incidents, which are not representative of the typical cybercrime experienced by smaller businesses and government institutions.

Third, estimates of victimisation rate from central reporting agencies (such as UK Report Fraud or US FBI IC3) face under-reporting bias: only a fraction of crimes are reported, making true victimisation rates impossible to determine directly.

The relevant victim population (e.g. adults over 18 in the US, or all registered businesses) can typically be obtained from census data or business registries with reasonable accuracy. However, scaling from a surveyed population to a broader one (e.g. UK to global, or businesses to all economic entities) introduces uncertainty, which we address in Section 4.

## Reviewing Commonly Cited Estimates

The estimates outlined in Table 2 have different strengths and weaknesses. In this section, we review the strengths and limitations of two estimates which are commonly cited for global cybercrime damages: inference from FBI Crime Complaint Center data and RAND's 2018 economic modeling-based estimates. We use this section to highlight the methodological problems we discussed in the previous section using real data.

### Inference from FBI Crime Complaint Center Data

**An Overview of the Original Data Source**

Many countries have centralized reporting systems which allow individuals and businesses to report cybercrime. The most prominent of these is the FBI's Crime Complaint Center (IC3), which is the USA's "primary destination for the public to report cyber-enabled crime and fraud as well as a key source for information on scams and cyber threats."[22] IC3 collects complaints from both individuals and from businesses. Cybercrime victims are "encouraged and often directed by law enforcement" to file cybercrime complaints to IC3.[23]

---

[21] Chainalysis, "The 2025 Crypto Crime Report," February 2025.
[22] FBI Internet Crime Complaint Center, "Internet Crime Report 2024," 6.
[23] FBI Internet Crime Complaint Center, "Internet Crime Report 2024," 3.

The IC3 collects reports of a wide variety of cybercrimes, ranging from fraud and scams against individuals to corporate data breach reports and hacking. In terms of the NAS taxonomy, this encompasses direct losses for both individuals and businesses under categories (1) and (2). In 2024, the IC3 received nearly 860,000 complaints, around 256,000 of which were associated with an actual loss of money. The average loss per complaint was around $19,000, and thus total reported losses were $16.6 billion.[24]

**A Description of Our Method Using the FBI IC3 Data**

The FBI IC3 data is not intended to be a direct, complete estimate of global cybercrime, despite often being used as such. Its limitations for that purpose include:

1. **Covers predominantly US victims.** The FBI IC3 collects reports of crimes predominantly from US nationals and uses the database to work with US local law enforcement agencies. While the system accepts international submissions, the vast majority of complaints are US-based. Approximately 15% of 2024 complaints listed a non-US complainant address,[25] though this reflects reported location rather than nationality. This means we need to find some way to extrapolate predominantly US damages to global damages.

2. **Suffers from under-reporting.** The head of the IC3 stated that the number of complaints only represented 10% to 12% of cybercrime victims in the US in 2016.[26] While the basis for this estimate is unclear, it aligns broadly with other estimates of cybercrime reporting rates. The UK National Crime Agency estimates a reporting rate of around 14%, while the reporting rate for businesses appears to be around 10% to 18%.[27]

3. **Only includes direct cost.** The IC3 data does not capture response and defense costs.[28] When filing a report, transaction information is requested to back up claims about losses, suggesting that only direct financial losses are included.

4. **The scope of covered costs is ambiguous.** While IC3 documentation indicates that remediation costs are excluded for ransomware incidents, substantial loss amounts are reported for business data breaches ($364M[29]). This is notable given the lack of an obvious wire transaction or direct financial transfer in such incidents.[30] It is also

---

[24] FBI Internet Crime Complaint Center, "Internet Crime Report 2024," 4.
[25] FBI Internet Crime Complaint Center, "Internet Crime Report 2024," 16.
[26] Al Baker, "An 'Iceberg' of Unseen Crimes: Many Cyber Offenses Go Unreported," *The New York Times*, February 5, 2018.
[27] National Crime Agency, "Fraud," n.d.; European Commission, "SMEs and Cybercrime," May 2022; See DSIT, "Cyber Security Breaches Survey 2024," April 9, 2024.
[28] FBI Internet Crime Complaint Center, "Internet Crime Report 2024," 10.
[29] FBI Internet Crime Complaint Center, "Internet Crime Report 2024," 10.
[30] It might be that these "direct losses" are extortion payments, though this is not mentioned in the FBI IC3 report. See e.g. recent data extortion campaign in Peter Ukhanov et al., "Oracle E-Business Suite Zero-Day Exploited in Widespread Extortion Campaign," Mandiant, October 9, 2025.

unclear whether all of the losses recorded in IC3 data ultimately represent permanent financial losses. For example, FBI sources indicate that the Recovery Asset Team is able to freeze or recover roughly 70% of funds in certain domestic business email compromise cases,[31] though this applies to only a subset of incidents (far smaller than the $2.7 billion in losses reported overall in 2022).[32]

In our simple model, we correct for the first two factors by making adjustments to the headline FBI IC3 loss estimates.

For a 'simple' estimate, we can scale the loss estimates to the world by assuming that cybercrime losses scale with GDP. Since world GDP is 3.8x US GDP,[33] we scale the losses by 3.8x. Secondly, to account for under-reporting, we assume a reporting rate of 10%, and assume that the average loss for unreported cybercrime is the same as for reported cybercrime. Table 3 below summarizes this calculation.

| Variable | Value | Description | Source |
| --- | --- | --- | --- |
| Aggregate losses (2024) | $16.6B | Hard-coded input | FBI IC3 |
| Under-reporting scale-up | 10X | Subjective assumption | Claim by former IC3 Director (Baker, 2018) |
| US to global scale-up | 3.8X | Subjective assumption | World GDP/US GDP |
| Inferred damages (Simple) | $630.8B | Calculation | |

**Table 3 | Inferring Cybercrime Damages.** A simple model inferring global cybercrime damages from FBI data.[34]

**Limitations**

Even making these adjustments, this estimate has several limitations that could bias it in either direction.

For example, the under-reporting scale-up may be too large and therefore creating an overestimate. The under-reporting rate assumed (10%) is at the lower end of other estimates. Moreover, the assumption that the average loss for unreported cybercrime is the same as for reported cybercrime, may not be correct. It is plausible that victims suffering larger losses are more likely to report the loss to IC3. Thus, the 10X "under-reporting scale-up factor" may be too high. If we assume a 90% reporting rate (i.e. minimal under-reporting, which seems extreme), the implied damages are around $70 billion. Second, scaling damages by GDP may not be appropriate. This assumes that wealthier countries present more lucrative targets; however, some countries may be targeted more than others for non-economic reasons.

---

[31] FBI Internet Crime Complaint Center, "IC3 Asset Recovery Team," n.d.
[32] FBI Internet Crime Complaint Center, "Internet Crime Report 2022."
[33] World Bank, "GDP (Current US$)," World Bank Group, 2024.
[34] The full calculations are in [Cybercrime] Incidence Rates and Lossess.

As we noted above, this estimate only includes direct costs, and so only captures a subset of cybercrime damages, excluding damages from response and defense.

Overall, we have low to moderate confidence in this estimate as an estimate of direct losses. A plausible range of losses that assumes that the percentage of cybercrimes that are ultimately reported to the FBI is between 10% and 90% is $70 billion to $630 billion.

## RAND Economic Modeling-Based Estimates

**An Overview of the Original Data Source**

The RAND Corporation published a 2018 report attempting to estimate global cybercrime costs through economic modeling.[35] The report employs two distinct modeling approaches across three models, producing estimates that range from hundreds of billions to tens of trillions of dollars. All methods aim to capture direct, response, and indirect costs to businesses across all industry sectors globally.

All three models seem to cover all cyber-attacks on the networks of economically productive sectors; this implies that this covers costs of category (1) of the NAS taxonomy ("Acts Targeted Against Machines, Data, or Systems") discussed in Section 2.

**A Description of the Methods Used**

**Models 1 and 2** share a common framework: they calculate the value that could be "successfully destroyed, stolen, or otherwise lost" across different cyber-attack types ("perils"), financial exposures (IP, capital assets, income), and industry sectors. Both models go sector-by-sector, estimating how much economic value could be destroyed by cyber-attacks, then add the downstream effects when one sector's disruption ripples through to others.

Model 1 derives both its exposure fractions (what fraction of each asset type is vulnerable) and peril impact rates (what fraction of vulnerable assets would actually be lost in an attack) from a 2016 analysis by Deloitte of approximately 50 Dutch companies, adapted and mapped to the RAND taxonomy.[36] Model 2 uses the same peril impact rates but estimates exposure fractions differently, using regression analysis of financial data from 4,447 publicly traded companies in SEC filings, relating R&D spending (as a proxy for IP), total assets, and net income to revenue across sectors.

---

[35] Paul Dreyer et al., "Estimating the Global Cost of Cyber Risk: Methodology and Examples," (Santa Monica, CA: RAND Corporation, 2018).
[36] B. Jacobs, J. Bulters, and M. van Wieren, "Modeling the Impact of Cyber Risk for Major Dutch Organizations," in Proceedings of The 15th European Conference on Cyber Warfare and Security, ed. by Robert Koch and Gabi Rodosek (Academic Conferences and publishing limited, 2016); Deloitte, "Cyber Value at Risk in the Netherlands," 2016.

**Model 3** takes a different approach to estimate the total "value at risk" in cyber-attacks using the Advisen database of approximately 900 cyber events with cost estimates from 2005-2015.[37] The approach then computes the fraction of revenue lost for a given company in a given incident; this fraction of revenue is then applied to the total output of a given sector, and scaled to a number of industry sectors. This process is done a number of times (what they call "bootstrap resampling") to create distribution of values at risk.

The resulting estimates are:

- **Model 1:** Direct costs of $275 billion; total costs of $799 billion
- **Model 2:** Direct costs of $3.2 trillion; total costs of $10.1 trillion
- **Model 3:** Direct costs of $6.6 trillion; total costs of $22.5 trillion

**Limitations and Biases**

All three models face significant methodological challenges that severely limit their reliability.

**Models 1 and 2** share a fundamental limitation: neither incorporates any probability of attacks occurring. In risk terminology (risk = likelihood × impact), these models estimate only the "impact" component while ignoring likelihood. They calculate what could be lost if all identified exposures were successfully attacked, not what is lost in practice. Model 1 relies on survey responses about cyber-vulnerable assets (exposure fractions) from just 50 Dutch companies. Model 2, while based on larger-scale data, attempts to adjust for the subjectivity by assuming that all assets are vulnerable: the report acknowledges it "includes all exposures, regardless of whether they could be affected by cyberattacks (e.g. a cornfield would be included)."[38] This explains why Model 2's estimates are over 10 times larger than Model 1's despite using the same framework.

**Model 3** suffers from severe unjustifiable probability and cost extrapolations. The method makes the implicit assumption that all businesses in a sector experience the same fraction of lost revenue as the business with a recorded incident in the Advisen database. However, the sample of 900 events is drawn from incidents severe enough to be recorded in the Advisen database – these represent atypically costly events rather than typical cyber damages.

Overall, we have very low confidence in these estimates as measures of actual realized cybercrime costs. This is only natural, since it wasn't the objective of the authors of the RAND report. All three models are better understood as theoretical maximum exposure analyses rather than empirical cost estimates. Interpreting $22.5 trillion as realized cybercrime damages would imply that costs are about 22% of World GDP, which is not reasonable. The range in estimates (2 orders of magnitude between the lowest and highest) illustrates the fundamental uncertainty in model-based approaches. These estimates may have value for scenario planning or understanding theoretical worst-case exposures, but should not be interpreted as estimates of current annual cybercrime damages.

---

[37] See section "Alternative Method for Directly Estimating Potential Economic Damage" for full details on this method.
[38] Paul Dreyer et al., 17,

## Our Composite Estimate of Global Cybercrime Damages

Our best estimate of global cybercrime costs combines three distinct estimation approaches, each targeting different categories of costs within our taxonomy. This composite method is necessary because different sources measure losses separately for different victim types and cost categories, and no single source provides comprehensive coverage of all cybercrime costs. Specifically, we estimate: (1) business and institutional direct and response costs, (2) business defense spending, and (3) individual direct losses. While this approach involves some overlap between categories – particularly where defense spending may encompass some response costs – it represents the most methodologically sound approach given available data sources. In the following subsections, we detail each component of our estimate.

### Businesses: Direct and Response Costs

**An Overview of the Original Data Source**

The UK Department for Science, Innovation and Technology (DSIT), in partnership with the Home Office, conducts an annual survey of businesses and charities to assess cybersecurity spending and breach experiences. The 2025 survey included 2,180 UK businesses (that have employees) via random probability telephone and online survey. It provides the most comprehensive publicly available data on business cybersecurity costs.[39]

The survey found that 43% of businesses reported experiencing some form of cyber security breach or attack in the last 12 months.[40] However, not all breaches result in material harm. Among businesses that experienced breaches, approximately 16% reported a "negative outcome" such as stolen money, corrupted software, temporary loss of access to files or networks, or other material impacts.[41]

The survey collects financial cost data for the most disruptive incident experienced by each organization. The DSIT categorizes these costs into four groups, though their taxonomy does not perfectly align with our own framework but it helpfully breaks down costs into sub-categories that we can then use to reconstruct: short-term direct costs, long-term direct costs, staff-time costs, and indirect costs.

According to the survey, the average total cost of the most disruptive breach or attack from the last 12 months, among those identifying a breach or attack with an outcome was £8,260.[42]

The DSIT values represent the financial cost of the most disruptive incidence – not the total annual cost across all incidents. Such a total number is not provided. However, the survey does separately ask about the frequency of any cyber-attacks. Among those identifying a breach in the last 12 months, 19% reported experiencing only one, 28% less than once a month, 23% once a month, and 29% weekly or more (Fig 4.5, DSIT, 2025). This implies an 'average' of ~20 attacks a year, again not all of which may have resulted in a "negative outcome" let alone "most disruptive".[43]

---

[39] DSIT, "Cyber Security Breaches Survey 2025," Section 1.3.
[40] DSIT, "Cyber Security Breaches Survey 2025," Section 4.2.
[41] DSIT, "Cyber Security Breaches Survey 2025," Section 4.8.
[42] DSIT, "Cyber Security Breaches Survey 2025," Tab 4.5b.

**Advantages of this Data Source**

Victimisation surveys represent one of the more methodologically sound approaches to estimating cybercrime costs, as they address the severe under-reporting problem that plagues aggregated complaint databases.[44] By randomly sampling organizations and directly asking about their experiences, these surveys can establish more accurate prevalence rates and average losses.

While several other business cybercrime victimisation surveys exist, including surveys from the Netherlands,[45] the European Commission,[46] and Accenture,[47] the DSIT survey offers several unique advantages. First, it provides transparency around survey design and question phrasing, publishing technical annexes and thorough explanations of their cost categories. Second, it reports losses as mean monetary amounts rather than just percentages or medians, enabling us to make more transparent inferences. Third, it surveys broadly across businesses of all sizes and sectors, making it more representative of the full business population. These methodological strengths make it our preferred source for estimating business cybercrime costs. In terms of the conceptual framework from Section 3.1, the DSIT survey provides estimates of all three key inputs: victimisation rate, average cost per incident, and victim population.

**Our Estimation Method Using the DSIT data**

The DSIT survey data does not directly provide a global estimate of cybercrime damages and explicitly cautions about extrapolating financial cost data, since margins of error are likely to be very wide.[48] However, since we are primarily interested in an approximate order-of-magnitude estimate of cybercrime damages, we still believe there is still value for us doing so. Nonetheless, we follow DSIT's suggested best practices of clearly labelling the extrapolation and calibrating the resulting estimate against other evidence as well.

To construct a global estimate, we first calculate the estimated 'most disruptive breach' costs experienced by the average surveyed UK businesses. This is done by multiplying the percentage of businesses experiencing breaches (43%), the percentage of those breaches resulting in negative outcomes (16%), and the average total cost of the 'most disruptive' breach of those affected (£8,260 total cost). Table 4 below shows it implies a point estimate of ~$726 per surveyed business.

---

[43] The ~20 attacks per year is derived from an approximation by the authors of this work as follows: 19% * 1 [only once] + 28% * 5 [less than once a month] + 23% * 12 [once a month] + 29% * 52 [weekly or more].
[44] Anderson et al.
[45] Central Bureau voor de Statistiek, "Cybersecuritymonitor 2023," June 28, 2024.
[46] European Commission, "SMEs and Cybercrime."
[47] Accenture Security, "The Cost of Cybercrime," Accenture, 2019.
[48] DSIT, "Cyber Security Breaches Survey 2025," Annex 1.7.

Second, we now try to extrapolate this average per business to the entire UK economy and then the world. As mentioned, this relies on making several critical assumptions:

- **Estimating Annual Cost:** As noted, the DSIT survey only estimated the cost of the "most disruptive" incidence, not the total of all incidents. To bound this we can consider two approaches. A conservative approach would be that the worst incidence is the only cost. An aggressive approach would be to assume each of the ~20 cyber-attacks experienced by the average business in 12 months is as bad as the "most disruptive" one. We think the truth is likely closer to the former than the latter since incidents are plausibly fairly right-tail distributed and so the 'worst incidence' carries a lot of the total damages. Thus, we make a subjective judgement of assuming the total cost is ~3X the "most disruptive incidence," though we remain highly uncertain about the true value.

- **Extrapolating to all UK businesses:** As per the DSIT survey, there are ~1.43M UK businesses with employees that such average costs can then be extrapolated too. The margins of error here are notably smaller than the other parameters.

- **Extrapolating to the UK economy:** The financial cost numbers taken from the DSIT survey only cover private-sector businesses. It does not include cybercrime damages accrued by government organizations, which in the UK includes hospitals and other services. Given ~44% of UK GDP is government expenditure,[49] we might employ GDP scaling and assume cybercrime costs scale proportionally with economic activity. This gives a scale-up factor of 1.8X.

- **Extrapolating to the world:** Similarly, to scale from UK to global costs, we might employ GDP scaling under the assumption that cybercrime costs scale proportionally with economic activity. Since world GDP is approximately 30X times larger than UK GDP,[50] our subjective judgement is centered on that, but we include uncertainty.

Table 4 summarizes the key assumptions and calculations underlying this estimate. We see that the total is ~$200B, of which ~$125B is what per our taxonomy may be considered 'remediation costs'.

---

[49] International Monetary Fund, "Government expenditure, percent of GD," 2025.
[50] World Bank, "GDP (Current US$)."

| Variable | Value | Description | Source |
|---|---|---|---|
| Percentage of businesses having a breach in last 12 months | 43% | Hard-coded input | DSIT Survey, 2025; Sec 4.2 |
| Percentage of businesses having negative outcome, of those affected | 16% | Hard-coded input | DSIT Survey, 2025; Fig 4.6 |
| Mean reported total cost of worst incident of those affected | £8,260 | Hard-coded input | DSIT Survey, 2025; Tab. 4.5b |
| Mean reported total direct cost of worst incident of those affected | £3,110 | Hard-coded input | DSIT Survey, 2025; Tab. 4.1b |
| *Of which remediation* | £5,150 | Calculation | |
| Average GBP to USD Exchange Rate in 2024 | 1.3x | Hard-coded input | IRS (2025) |
| Mean reported total direct cost of worst incident (USD 2024) | $726 | Calculation | [=43%*16%*£8,260*1.28] |
| Worst incidence to Total incidences | 3.0x | Subjective assumption | DSIT Survey, 2025; Fig 4.6 |
| Total business population of the UK | 1.4M | Hard-coded input | DBT, 2024 |
| Private Sector to Economy scale-up | 1.8x | Subjective assumption | UK GDP / (1- UK Gov. Exp.) |
| UK to Global scale-up | 32x | Subjective assumption | World GDP/UK GDP |
| **Inferred business damages** | **$200B** | **Calculation** | [=$700*3x* 1.4M*1.8x*31x] |
| *Of which remediation* | $125B | | |

**Table 4 | Business Cybercrime Damage.** A simple model inferring global business cybercrime damages from DSIT data

**Limitations**

This estimate faces several limitations that could bias the results in either direction. First, as noted, many of the subjective assumptions are uncertain. Including these is inherently required for arriving at an aggregate global estimate of cybercrime – and we have attempted to be very transparent about our assumptions. Nonetheless, future work may benefit from surveying a larger group of subject matter experts to better estimate these parameters.

Second, GDP scaling assumes that cybercrime victimisation rates and costs scale proportionally with economic activity across all countries. This assumption may not hold for several reasons. Different sectors of the economy may experience different victimisation rates and losses than businesses, yet we apply business victimisation rates and average losses uniformly. For example, hospitals may be a more or less frequent target than average. Moreover, countries vary substantially in many factors, such as different levels of contributions of various sectors to the total economy, cyber defense capabilities, regulatory environments, and exposure to cyber threats. For instance, nations more involved in geopolitical conflicts (such as the United States, China, or European nations in the context of the Russia-Ukraine war) may face higher rates of state-sponsored attacks that could disproportionately increase costs. Future work may want to do such robustness checks.

Third, we note it is possible that the DSIT survey's sampling strategy might result in underestimating the average financial cost of incidents. We note that the DSIT survey defines a "Large business" as having 250 or more employees and surveyed <80 of these for its financial cost data. However, this definition of "large business" hides a very long tail: some large businesses have over 100,000 employees.[51] Thus, if, as appears reasonable to us, the costs of cybercrime scales with size, the average recorded in the survey might be heavily determined if one of these mega-employers is included in the sample or not.

Fourth, the DSIT cost categories do not align perfectly with our taxonomy. The survey's "indirect costs" category includes items such as staff time when they could not do their jobs,[52] which we classify as response costs rather than indirect costs. Conversely, some of their "long-term direct costs" overlap with our response cost category. Additionally, the DSIT survey distinguishes between "cyber security breaches and attacks" and "cyber crimes"[53] when, for our purposes, this distinction has limited practical impact because we focus exclusively on breaches that resulted in material negative outcomes.

Despite these limitations, we believe this presents the best attempt at quantifying business cybercrime costs globally in spite of some inherent limitations. The DSIT survey's rigorous methodology, large sample size, and transparent reporting make it substantially more reliable to build an estimate from – especially when comparing to alternative approaches such as analyzing stock price movements following breach disclosure or modeling based on small samples of high-profile incidents.

---

[51] See e.g. Tesco, "Sustainability Factsheet 2023/24."
[52] DSIT, "Cyber Security Breaches Survey 2025," Section 4.6.
[53] DSIT, "Cyber Security Breaches Survey 2025," Section 6.2.

## Defense Costs

Several market research firms have published estimates of global spending on cybersecurity or information security, which we consider to be an estimate of global defense costs. These figures are drawn from proprietary reports that are prohibitively expensive to purchase, so we rely on publicly available summaries and cannot verify the underlying methodologies:

- "Worldwide end-user spending on information security … $193 billion in 2024"[54]
- "global cybersecurity IT investment reached $244.4 billion in 2024"[55]
- $154.6 billion in 2024[56]

An approximate midpoint gives us an estimate that global spending on cybersecurity is about $200 billion a year. However, this figure represents the total revenues of cybersecurity firms globally, encompassing both defensive spending and incident response (remediation) spending. Our previous estimate in [Section 4.1](#) computed global business direct and remediation spending, meaning that there's a partial overlap between these two estimates. Because we do not know how much of the $200 billion of global cybersecurity spending is spent on defense versus incident response by end users, we assume that it's simply half each. Therefore, our best estimate of global defense spending is **$100 billion** per year.

## Individual Cybercrime Costs

**An Overview of the Original Data Source**

Breen et al. conducted a large-scale victimisation survey of approximately 10,000 American adults (18 years and older) in the summer of 2020.[57] The survey assessed experiences with six specific types of cybercrime: non-delivery scams, non-payment fraud, extortion (including ransomware), overpayment scams, advanced fee fraud, and credit card or banking fraud. Unlike many cybercrime surveys that rely on vague or technical terminology, this study used plain language descriptions to ensure respondents understood what types of incidents to report.

---

[54] Gartner, "Gartner Forecasts Worldwide End-User Spending on Information Security to Total $213 Billion in 2025," July 29, 2025.
[55] IDC, "China Cybersecurity Market to Jump from US$11.2B (2024) to US$17.8B (2029) at 9.7% CAGR," Cyber Security Mew, August 25, 2025.
[56] Merrit Maxim, "Global Cybersecurity Spending To Exceed $300B By 2029," Forrester, October 3, 2025.
[57] Breen et al., "A Large-Scale Measurement of Cybercrime Against Individuals," *CHI '22: Proceedings of the 2022 CHI Conference on Human Factors in Computing Systems*, no. 122, (2022): 1–41.

For each crime type, respondents who reported being victimised were asked to provide the amount of money they lost. The study reports victimisation rates for each crime category as well as the distribution of losses, including median values and at the 10th and 90th percentiles. In the code release for their paper, the authors also report mean losses for each crime type. These distributions are notably right-skewed,[58] indicating that while most victims experience modest losses, a small subset experiences substantially larger losses. It also noted that "Based on the most recent FBI crime report statistics, the six cybercrimes we study represent 30% of cybercrimes against individuals."

The study's primary objective was to estimate prevalence rates and typical losses for American victims, making it well-suited for generating national and, with appropriate scaling, global cost estimates. The large sample size is particularly important given that cybercrime victimisation is a relatively rare event, requiring substantial sampling to achieve statistically meaningful results.

**Why We Prioritize It**

As mentioned before, a major advantage of victimisation surveys over other methods like aggregate complaint databases – such as the FBI's IC3 – is that they help to address the severe under-reporting problem. Several other individual cybercrime victimisation surveys targeting individuals exist, including surveys from Australia,[59] the Netherlands,[60] the US Federal Trade Commission,[61] and various industry-sponsored surveys. However, the Breen et al. survey offers several methodological advantages that make it our preferred source.

First, the sample size of 10,000 respondents provides sufficient statistical power to detect relatively rare victimisation events. This is substantially larger than many other academic surveys and reduces sampling uncertainty. Second, the survey design uses clear, accessible language to describe crime types rather than technical terminology, reducing the risk of misclassification by respondents. Third, the authors provide clear definitions and quantities for losses, allowing us to understand what drives the data. Fourth, this survey covers several cybercrimes against individuals directly, as opposed to some other surveys that cover, for example, only identity theft. In fact, Breen et al.'s terminology map onto the FBI's IC3 definitions, allowing us to do an additional sense check of our results. In terms of the conceptual framework from Section 3.1, this survey provides estimates of all three key inputs: victimisation rates and average costs per incident for each crime type, with victim population drawn from US census data.

---

[58] See Figure 2 in Breen et al.
[59] Isabella Voce and Anthony Morgan, "Cybercrime in Australia 2024," Australian Institute of Criminology, 2025.
[60] Central Bureau voor de Statistiek, "Online Veiligheid en Criminaliteit 2024," April 15, 2025.
[61] Keith B. Anderson, "Mass-Market Consumer Fraud in the United States: A 2017 Update," Federal Trade Commission, October 2019.

The Australian survey,[62] while comprehensive, suffers from a potentially self-selected sample and reports mean losses "after recoveries" based on only 40 respondents – far too small a sample for reliable extrapolation. Additionally, the survey questions appear to conflate personal and work-related cybercrime experiences, making it unclear whether reported costs represent individual or business losses.[63] The Netherlands survey[64] does not collect data on monetary losses, making it unsuitable for cost estimation despite its value for measuring prevalence.

**A Description of Our Method Using the Breen et al. 2022 Data**

To estimate global individual cybercrime costs from the Breen et al. survey data, we combine its data with several additional assumptions.

First, we combine the prevalence rate and mean reported loss for each crime type to estimate the expected loss per person per year. This value is $49.23. See Appendix A for details.

Second, we must account for the fact that these six crime types represent only a subset of all cybercrime against individuals. In the original work, Breen et al. calculate this by looking at what fraction of total individual cybercrime incidents these six types represented in the FBI IC3 complaint data, suggesting it is ~30%. Using the IC3 data for 2024 gives us a similar estimate when using prevalence as the weighting. We follow the authors' original methodology of using prevalence weighting but caveat that the correct weighting may be by dollar amount, as other areas of individual cybercrime not represented in the survey have grown a lot since 2020, especially via cryptocurrency and investment fraud.

Third, having now estimated the average total individual cybercrime loss, we can now scale this up first to the US and then the global cost. We do this by noting that there are 260M adult individuals in the US (as per the original Breen paper). Then, we scale from US to global costs using GDP ratios. Table 5 below summarizes the key assumptions and calculations underlying this estimate.

---

[62] Voce and Morgan.
[63] Voce and Morgan, 3.
[64] Central Bureau voor de Statistiek, "Online Veiligheid en Criminaliteit 2024."

| Variable | Value | Description | Source |
|---|---|---|---|
| Average estimated individual loss due to 6 tested cyber crimes | $49 | Calculation | Breen et al., 2022; Tab 5.; See Appendix A |
| Inflation adjustment ($2020 to $2024) | 1.25x | Hard-coded input | Bureau of Labor Statistics |
| Fraction of all cybercrime covered in survey, as of 2024 | 14% | Subjective assumption | IC3, 2024; see Appendix C |
| Total adult population of the US | 267M | Hard-coded input | US Census Data |
| US to global scale-up | 3.8x | Subjective assumption | World GDP/US GDP |
| **Inferred consumer damages** | **$240B** | Calculation | [=$49*1.25/30%*267M*3.8x] |

**Table 5** | **Individual Cybercrime Damages.** A simple model inferring global individual cybercrime damages from Breen et al. data.

**Limitations**

This estimate faces several important limitations that warrant careful consideration. First, the underlying survey was conducted in 2020, and the cybercrime landscape has evolved substantially since then. The dramatic growth in cryptocurrency-based investment scams and romance fraud operations, particularly those run from trafficking operations in Southeast Asia (see Appendix B for details on this topic), represents a significant shift in the fraud ecosystem that may not be fully captured by extrapolating 2020 victimisation rates. Our adjustment for this is only approximate.

Second, the calculation of what fraction of cybercrime the six surveyed types represent introduces considerable uncertainty. This fraction is calculated using FBI IC3 complaint data, which itself suffers from under-reporting and may not reflect the true distribution of cybercrime types. If certain crime types are systematically more or less likely to be reported than others, this could bias our extrapolation. The dramatic decline from 30% coverage in 2019 to 14% coverage in 2024 illustrates how rapidly the cybercrime landscape can shift. Future surveys may want to more directly ask about more types of cybercrime, instead of having to rely on our current extrapolation.

Third, as discussed in Section 4.1.1, GDP scaling assumes that cybercrime costs scale proportionally with economic activity, which may not hold across countries with varying levels of cyber defense capabilities, consumer protections, and cultural factors affecting victimisation and reporting rates.

Nonetheless, this estimate represents a good faith attempt at quantifying individual cybercrime costs globally in spite of some inherent limitations.

### Final Estimate

We estimate that individual direct losses to cybercrime are on the order of **$200 billion** yearly using the source we deem most reliable, the survey presented in Breen et al.[65] We estimate that the business direct and remediation yearly costs are in the range of **$200 billion**, inferring this value from the UK survey of cybersecurity breaches.[66] (DSIT, 2024). Combining these two estimates yields an estimate of global direct and remediation business and individual cybercrime costs of **~$400 billion** per year.

Our best estimate of additional global cybersecurity spending not captured in these above estimates (specifically on preventative technologies) is **~$100 billion a year**. If we include global defense spending, this estimate moves up to **~$500 billion per year**. We remain highly uncertain about the true value and therefore report our 90th percentile confidence interval as **$100 billion to $1 trillion per year**.

### Other Estimates

Numerical values of the other sources we considered, which are less-frequently cited, can be found in this spreadsheet. Note that many of these estimates are experimental and we expect there to be many limitations and incorrect assumptions.

---

[65] Breen et al.
[66] DSIT, "Cyber Security Breaches Survey 2024."

## Discussion

Our primary aim in this report was to better understand the approximate size of annual global cybercrime damages and be transparent about the underlying assumptions. This is a necessary first step in assessing the cybercrime risks posed by future AI capabilities.

### Estimate Limitations

In establishing our estimate we attempted to be very transparent in our methods. We note several common limitations affect our estimates:

**Observability limitations.** Our estimates may undercount total cybercrime costs for several reasons. First, state-sponsored cyber operations are often designed to avoid detection and attribution. For instance, the Volt Typhoon campaign maintained network access for up to five years in some cases before discovery.[67] This would imply that there are a number of ongoing campaigns that companies are themselves unaware of that will ultimately cause damages. Second, our heavy reliance on GDP scaling from surveys conducted primarily in Western nations (UK, US) fails to capture important variations across regions. Countries with less digitized economies, different regulatory environments, or varying exposure to geopolitical cyber conflict may experience substantially different victimisation rates and costs.[68]

**Unquantified costs.** Several categories of costs fall outside our measurement framework. Examples of unquantified costs include:

- Psychological trauma and stress experienced by fraud and investment scam victims

- Human suffering of people trafficked to perform fraud and investment scams (see [Appendix B](#) for details)

- Erosion of public trust in digital systems and online commerce, potentially constraining economic growth

- National security implications of persistent compromise of critical infrastructure and government institutions

While these costs are difficult to quantify in monetary terms, they represent genuine social harm that our estimates do not capture. They may well exceed the direct costs estimated in this report.

---

[67] Cybersecurity and Infrastructure Security Agency, "PRC State-Sponsored Actors Compromise and Maintain Persistent Access to U.S. Critical Infrastructure," February 7, 2024.
[68] For example, Kaspersky, an incident response firm based in Russia, finds that the Russia-complex of countries experience ransomware less frequently than the global average, likely due to many ransomware operators being based in Russia. See Assolini et al, "State of Ransomware in 2025," Kaspersky, May 7, 2025.

**Measurement uncertainty.** As discussed throughout Section 3, all available measurement methods face substantial limitations – from under-reporting in complaint databases to sampling uncertainty in victimisation surveys. The wide confidence intervals in our final estimates reflect this fundamental measurement challenge.

Future work improving cybercrime cost estimation should prioritize: developing more frequent and harmonized victimisation surveys across multiple countries, creating alternative indicators for detecting shifts in cybercrime patterns that don't rely solely on aggregate cost estimates, and establishing methodologies for quantifying the broader social costs of cybercrime beyond direct financial losses.

While limitations constrain the precision of our estimates, we nonetheless believe our work is helpful to decision makers. Prior to this work, estimates of global cybercrime damages ranged from tens of billions to tens of trillions of dollars, with little systematic evaluation of underlying methodologies and explanations of rationale. By critically examining 27 different sources, establishing a clear taxonomy of costs, and prioritizing the most methodologically sound approaches, we narrow this range considerably. Our estimate of ~$500 billion in annual cost (though still uncertain) provides a more solid foundation for evaluating cybercrime risk, including transparent assumptions for future work to expand upon.

## Relevance to AI Policy

**Aggregate cybercrime estimates are insufficient for detecting AI-enabled uplift or setting risk thresholds.** The uncertainties in global cybercrime damage estimates are simply too large to detect incremental changes caused by AI. There is no single data source that accurately captures total damages due to pervasive under-reporting, evolving crime definitions, and measurement inconsistencies across jurisdictions.

Consider again the framework presented in Figure 3, which conceptualizes total damages as the sum of products of average cost, victimisation rate, and victim population across crime types. Frontier AI developments could independently affect any of these factors. AI might increase victimisation rates through more automated targeting of existing attack chains. It might increase average damages per incident – for example, by enabling adversaries to more efficiently identify high-value assets during data exfiltration, as suggested by recent report of a threat actor using Claude to "independently query databases and systems, extract data, parse results to identify proprietary information, and categorize findings by intelligence value."[69] Or AI might expand the summation entirely by enabling novel attack surfaces that create additional damage categories (e.g. prompt injection attacks on AI-enabled services.[70]

---

[69] Anthropic, "Disrupting the first reported AI-orchestrated cyber espionage campaign," November 13, 2025.
[70] Brave, "Agentic Browser Security: Indirect Prompt Injection in Perplexity Comet," August 20, 2025.

Given these measurement challenges, future work should focus on developing alternative indicators that could collectively signal meaningful shifts. Promising approaches include: monitoring for AI-generated code signatures in malware samples, tracking criminal forum discussions of AI tool adoption, observing labor market changes in fraud operations (see [Appendix B](#)), and conducting periodic capability evaluations across representative attack chains. We propose that an alternative assessment approach could involve constructing a representative set of cybercrimes, tracking evidence across those crimes (safety test results, real-world incidents, security telemetry), and convening expert panels to estimate whether cumulative uplift crosses meaningful thresholds.

Given these measurement challenges, future work should focus on developing alternative indicators that could collectively signal meaningful shifts. We propose an alternative approach to assessing AI-enabled cybercrime risk that does not rely on detecting changes in aggregate damage estimates. This approach would involve the following steps:

1. Identify a set of attack chains spanning the major categories of cyber-dependent and cyber-enabled crime, weighted by their contribution to overall damages.

2. For each category, establish measurable signals that could indicate AI adoption or capability uplift, such as indicators of AI-generated code in malware samples, changes in labor dynamics within fraud operations (see [Appendix B](#)), or shifts in attack sophistication observable through security telemetry.

3. Establish a standing panel of cybersecurity practitioners, researchers, and economists who periodically synthesise evidence across these indicators to estimate whether cumulative uplift has crossed policy-relevant thresholds (e.g. a 20% increase in damages for a given crime category).

**Cybercrime represents substantial economic harm and triggers regulatory obligations, even if it does not constitute a "catastrophic risk."** Frontier safety frameworks and similar policies typically focus on catastrophic or existential risks characterized by instantaneous, irreversible harm. Cybercrime does not fit this pattern. It causes cumulative harm through many distributed incidents rather than a single catastrophic event.

However, this framing may create a blind spot. Our estimates suggest that even under our conservative assumptions, annual cybercrime damages are on the order of $500 billion a year, meaning that even a modest increase in damages due to AI usage may exceed the $100 billion damages threshold. Whether cumulative annual harm of this magnitude warrants similar attention to instantaneous catastrophic risks is a question that current frameworks do not clearly address.

Regardless of how one resolves this conceptual question, AI developers may face concrete obligations. The EU AI Act's Code of Practice explicitly identifies "enabling large-scale sophisticated cyber-attacks" as a systemic risk requiring assessment and mitigation.[71] Notably, this language does not require a single catastrophic event. It could reasonably encompass scenarios where AI enables a large-scale increase in smaller-impact attacks that collectively pose systemic risk. Compliance with these requirements does not depend on whether cybercrime qualifies as "catastrophic" under other individual company frontier safety frameworks.

Future work should clarify how cybercrime risk fits within the broader landscape of AI safety concerns and establish appropriate management approaches even where it falls outside traditional catastrophic risk definitions.

**Defensive adaptation is real but unevenly distributed, complicating net impact assessments.** Cybersecurity differs from many other risk domains because AI can substantially enhance defensive capabilities, not just offensive ones. Just as criminal actors face economic incentives to adopt AI tools, cybersecurity firms face strong market incentives to develop AI-powered defensive solutions. Any assessment of AI's net impact on the cybercrime landscape must account for these defensive improvements.

Defensive adaptation is unlikely to be uniform. Cybersecurity spending is concentrated among large enterprises and wealthy nations. Small and medium businesses, under-resourced government agencies, healthcare systems, and organizations in developing countries may not benefit equally from AI-enhanced defenses. Future modeling should account not only for aggregate defensive improvements but also for the distribution of those improvements across potential victims. If AI primarily benefits well-resourced defenders while enabling attacks on under-protected targets, the net effect could be harmful even if aggregate defensive capabilities improve.

**Under our estimates, a 20% increase in cybercrime damages would reach commonly cited catastrophic thresholds.** Our central estimate of approximately $500 billion in annual damages implies that only 20% uplift would be needed to cross the $100 billion incremental damage threshold that some frameworks identify as catastrophic. This is a relatively modest increase that could plausibly occur through efficiency gains in existing attack methods without requiring qualitatively new AI capabilities. The $100 billion threshold was originally motivated by the safety frameworks of leading AI companies, though some have since moved away from explicit quantitative thresholds.[72] Lower thresholds are also policy-relevant: a $1 billion damage threshold is used by at least one AI company and in California's SB53 legislation.[73] The question of where to set such thresholds remains under-researched.[74] Under our central estimate, reaching $10 billion in AI-driven incremental damages would require only ~2% uplift, and reaching $1 billion would require only ~0.2% uplift.

---

[71] European Union, Appendix 1.4.
[72] OpenAI's Preparedness Framework uses "hundreds of billions" as an explicit quantitative threshold for catastrophic harms (OpenAI, 2023, p. 2; retained in OpenAI, 2025, p. 2). Anthropic's original Responsible Scaling Policy similarly referenced this level (Anthropic, 2023, p. 1), but no longer provides an explicit quantitative threshold (Anthropic, 2025).

Moreover, this calculation likely understates the concern. As discussed above, our estimates exclude several categories of harm, such as psychological trauma, erosion of trust in digital systems, and national security implications. If these unquantified harms were included, the baseline would be higher and the uplift required to reach any given threshold would be correspondingly smaller.

This finding underscores the need for developing better measurement approaches. If AI-enabled uplift of even modest magnitude could cross policy-relevant thresholds, the current inability to detect such changes represents a significant gap in our capacity to govern AI-related cyber risks.

---

In conclusion, this report narrows plausible estimates from spans of multiple orders of magnitude to a more transparent baseline. While limitations remain, our systematic methodology provides a foundation for more informed decision-making about AI-related cybercrime risks. Additionally, we have pointed to several further useful directions of research.

---

[73] xAI and California's SB53 use a $1 billion threshold. xAI, "xAI Risk Management Framework," (August 20, 2025), pg 3. California Legislature, Senate Bill 53, "Artificial intelligence models: large developers," 2025–2026 Regular Session, approved September 29, 2025.
[74] On the broader question of how and where companies should set risk thresholds, see: Leonie Koessler, Jonas Schuett, and Markus Anderljung, "Risk Thresholds for Frontier AI," (June 20, 2024).

## About the Authors

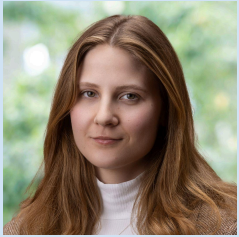

**Kamile Lukosiute** ✉ 𝕏 in g
Research Scholar, GovAI

Kamile's research focuses on threat modeling of cyber adversary misuse of advanced AI systems. Previously, she worked at Cisco Security as an AI Security Researcher and at Anthropic as a Resident in AI alignment research. She holds an MS in Physics from the University of Amsterdam.

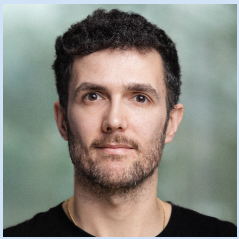

**John Halstead** ✉ in g
Research Fellow, GovAI

John works on threat modeling, with a particular focus on CBRN and cyber misuse risk. Before joining GovAI he worked at the Forethought Foundation and Founders Pledge. He holds a DPhil in Political Philosophy from the University of Oxford.

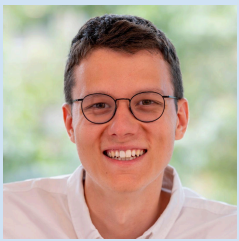

**Luca Righetti** ✉ 𝕏 in g
Senior Research Fellow, GovAI

Luca is a Senior Research Fellow at the GovAI, where he leads a team to investigate national security risks from advanced AI systems. He previously worked at Open Philanthropy, the UK Office for AI, and the University of Oxford's Future of Humanity Institute.

## Acknowledgements

We would like to thank Adam Swanda, Asher Brass, Tom Johansmeyer, Matt Malone, Nicholas Butts, Connor Aidan Stewart Hunter, Bhuvana Sudarshan, Matthew van der Merwe, and Alan Chan for reviewing an earlier draft of this work.

## About GovAI

GovAI is a 501c(3) non-profit organisation. Our mission is to help decision-makers navigate the transition to a world with advanced AI, by producing rigorous research and fostering talent. Researchers at GovAI work on a wide range of topics, with a particular emphasis on the security implications of frontier AI.

# Appendix

## Appendix A | Estimating Total Average Loss from Breen et al. 2022

Breen et al. (2022) report prevalence values for six types of cybercrime in Table 5 of the original paper.[75] The mean dollar losses for each crime type are published in the authors' code release. We calculate the expected loss per person as:

$$E[loss] \;=\; \sum_i prevalence_i \times mean\,loss_i$$

| Crime Type | Prevalence | Mean Loss | Contribution |
|---|---|---|---|
| Bank/CC (lost money) | 1.082% | $3,545.95 | $38.37 |
| Non-Delivery | 3.205% | $116.66 | $3.74 |
| Advanced Fee | 0.280% | $1,659.49 | $4.65 |
| Non-Payment | 0.344% | $419.07 | $1.44 |
| Extortion | 0.116% | $759.49 | $0.88 |
| Overpayment | 0.052% | $376.29 | $0.20 |
| Total | | | $49.27 |

**Table A1 | Mean losses and prevalence for each of the six crimes reported in Breen et al.** Mean losses are from the authors' codebase; prevalence values are from Table 5 of Breen et al. (2022).

Using the published mean loss values combined with the prevalence, the **expected loss per year per person** due to the six studied cybercrimes in the **U.S. is $49.27.**

## Appendix B | Scam Center Criminal Revenues

As discussed in the main report, not all social costs of cybercrime are captured by economic damages numbers. One significant cost is the humanitarian problem of people in forced labor at scam operations. The relationship between AI automation and this particular harm is complex and nuanced: if AI systems become cheaper and more effective at conducting scams than coerced human labor, demand for trafficked workers could decline; if it significantly increases their productivity, the number of trafficked workers could increase. In this section, we discuss the scale of the problem and our estimates of the revenues generated by these workers as a backup to our primary estimates presented in the main text.

---

[75] Breen et al.

Human trafficking to countries in South East Asia as well as Dubai to forcibly work as scammers is a well-documented recent problem.[76] Lured by promises of legitimate work, victims from China, India, Africa, as well as locals from the area, are transported to compounds in Laos, Cambodia, Myanmar, and to a lesser extent Dubai, Thailand, Philippines, and China, and forced to work scamming people online, pretending "to be an attractive woman, targeting men who seemed affluent."[77] The accounts of people who have managed to escape are harrowing, with people describing extreme violence, and victims going through extreme lengths to escape, including risking death.[78] The scale of the problem seems very large: UN estimates indicate that around 120,000 people in Myanmar and around 100,000 people in Cambodia may be "held in situations where they are forced to carry out online scams."[79] Estimates from Chinese state media seem to also place similar estimates on the number of people involved in these scams in the region.[80]

As people are rescued from these operations or escape, they often give interviews, citing daily or monthly quota numbers that, if not met, result in violent punishments, or numbers for ransom that, if paid, would release them from bondage.[81] Using these reported numbers, several organizations have attempted to estimate the total global earnings of the crime syndicates that are ultimately benefitting from the scamming. The revenue of scamming operations should, in principle, be equal to the losses that are experienced by the victims (minus costs of laundering, such as cuts taken by crypto mixers) and therefore this estimate is useful for our purposes for understanding the scale of the fraud ecosystem. The first estimate comes from a UN report that claims that the revenue for "criminal networks engaged in cyber-enabled fraud generate between US $27.4 and $36.5 billion annually" using the "estimated labor force in scam centers in 10 Southeast Asian countries" and "the average amount of proceeds generated (or stolen) by each individual."[82] Unfortunately, this is the extent of the methodology provided by the UN for this estimate. The United States Institute of Peace performs a similar estimate for global annual revenue, combining estimates of global workforce with a $350 daily revenue ($10,500 per month) per person to arrive at an estimate of $63.9B annually.[83]

---

[76] United Nations Office on Drugs and Crime (UNODC), "Transnational Organized Crime and the Convergence of Cyber-Enabled Fraud, Underground Banking and Technological Innovation in Southeast Asia: A Shifting Threat Landscape," October 2024; Charlie Campbell, "Poppies to Pig-Butchering: Inside the Golden Triangle's Criminal Reboot," *Time*, March 21, 2024; Alastair McCready, "Inside the Chinese-run crime hubs of Myanmar that are conning the world: 'we can kill you here,'" *SCMP*, July 22, 2023; United States Institute of Peace, "Transnational Crime in Southeast Asia," May 2024; Human Research Consultancy, "Uncovering the Spread of Human Trafficking for Online Fraud into Laos and Dubai," USAID, July 2024.
[77] Cait Kelly, "'I couldn't escape': the people trafficked into call centres and forced to scam Australians," *The Guardian*, April 1, 2023.
[78] Poppy McPherson and Napat Wesshasartar, "Thai teen says he threw himself out of a window to escape Cambodia's brutal scam farms," *Reuters*, June 27, 2025.
[79] UN OHCHR, "Online Scam Operations And Trafficking Into Forced Criminality In Southeast Asia: Recommendations For A Human Rights Response," 2023.
[80] United States Institute of Peace, "Transnational Crime in Southeast Asia."
[81] Kelly; McPherson and Wesshasartar.
[82] UNODC, "Transnational Organized Crime and the Convergence of Cyber-Enabled Fraud, Underground Banking and Technological Innovation in Southeast Asia: A Shifting Threat Landscape."

We could not identify a source for the $350 daily revenue estimate and therefore we are compelled to compute our own estimate of global criminal revenues. There are relatively credible estimates of the amount of people forcibly working in scam centers around the world. We use the estimates from the United States Institute of Peace[84] and use 370,000 people worldwide as a conservative estimate, as it is based on counts of reputable reports and 500,000 as an upper bound following their reasoning about post-COVID-19 work pivots of Chinese casino employees.

For estimates of revenue, we use two distinct numbers. First, we find a reported range of $25,000 to $50,000 for the expected monthly quota; however, these may be over-estimates as one victim report states that she was "one of the boss's highest earners, clearing more than $30,000 every month."[85] Our estimate of the global scammer revenues using this method provide a range of $111 – $300B in yearly revenues for global organized crime cyber-enabled fraud. However, this seems to be an extremely large over-estimate. These numbers are large because the reports of quotas are not necessarily the reports of real earnings; it's not clear how often workers actually hit the quota or if its true purpose is as a scare tactic, nor is it known what the average tenure of a scam worker is.

We may be able to get a lower bound estimate using a different method. As part of this ecosystem of human trafficking for scamming, people would be offered release in exchange for payment (ransom), or sometimes, they would be sold to rival gangs for labor. The crime boss should not care either way if the total value from an individual scammer is extracted via ransom, payment, or scamming labor; therefore, the ransom or sale price for a scammer should be roughly equal to the lifetime value generated for the crime boss. Assuming that the average tenure of a scammer is about a year (an assumption with little backing), this allows us to estimate the expected value that crime bosses should be expecting to generate per year using these ransom values, times the number of scammers. There is a wide range of reported ransoms or sale prices of scammers, depending on their effectiveness; this means our estimated range of global earnings is wide with $728.9M to $35.0B in expected value generated.

This is an incredibly incomplete estimate of the scale of the fraud ecosystem, as it is entirely unclear what fraction of the total global scammer population the people being held against their will in scam compounds represent. However, it is an illustrative side-analysis of how lucrative this particular crime area is. Our intuition is that the true range of global scam worker earnings is somewhere between the high end of the ransom estimate ($35B) and the low end of the scammer revenues ($111B).

---

[83] United States Institute of Peace, "Transnational Crime in Southeast Asia."
[84] United States Institute of Peace, "Transnational Crime in Southeast Asia."
[85] Campbell.

GLOBAL CYBERCRIME DAMAGES  |  TECHNICAL REPORTand Security.
https://books.google.co.uk/books?dq=economic+impact+of+cyber+risk&id=ijaeDAAAQBAJ

Johansmeyer, Tom. (2025, May 22). *Surprising stats: the worst economic losses from cyber catastrophes*. The Loop.
https://theloop.ecpr.eu/surprising-stats-the-worst-economic-losses-from-cyber-catastrophes/

Kelly, C. (2023, April 1). *'I couldn't escape': the people trafficked into call centres and forced to scam Australians*. The Guardian.
https://www.theguardian.com/law/2023/apr/02/i-couldnt-escape-the-people-trafficked-into-call-centres-and-forced-to-scam-australians

Khotari, S.P. & Warner J.B. (2006). *Econometrics of Event Studies*. In Handbook of Corporate Finance: Empirical Corporate Finance.
https://www.bu.edu/econ/files/2011/01/KothariWarner2.pdf

Koessler, Leonie, Jonas Schuett, and Markus Anderljung. "Risk Thresholds for Frontier AI." arXiv:2406.14713 (June 20, 2024). https://arxiv.org/abs/2406.14713.

Kosinski, M. (2024, July 30). *What is a data breach?* IBM.
https://www.ibm.com/think/topics/data-breach

Langton, L., et al. (2012, August). *Victimizations Not Reported to the Police, 2006-2010*. National Crime Victimization Survey (NCVS). U.S. Department of Justice.
https://bjs.ojp.gov/content/pub/pdf/vnrp0610.pdf

Leyden, J. (2002, November 21). *Why is mi2g so unpopular?* The Register.
https://www.theregister.com/2002/11/21/why_is_mi2g_so_unpopular/

McCready, A. & Mendelson, A. (2023, July 22). *Inside the Chinese-run crime hubs of Myanmar that are conning the world: 'we can kill you here'*. South China Morning Post (SCMP).
https://www.scmp.com/week-asia/politics/article/3228543/inside-chinese-run-crime-hubs-myanmar-are-conning-world-we-can-kill-you-here

Meta. (2025). *Frontier AI Framework*. Version 1.1.
https://ai.meta.com/static-resource/meta-frontier-ai-framework/?utm_source=newsroom&utm_medium=web&utm_content=Frontier_AI_Framework_PDF&utm_campaign=Our_Approach_to_Frontier_AI_blog

Meurs, T. (2024). *Double-Extortion Ransomware: A Study of Cybercriminal Profit, Effort, and Risk*. Dissertation, University of Twente.
https://ris.utwente.nl/ws/portalfiles/portal/475998586/Final_PHD_Dissertation_jan25.pdf

Meurs, T., Junger, M., Cruyff, M. & van der Heijden, P.G.M. (2025, July 29). *Estimating The Number Of Ransomware Attacks*. Journal of Quantitative Criminology.
https://link.springer.com/article/10.1007/s10940-025-09625-7
GovAI  |  45

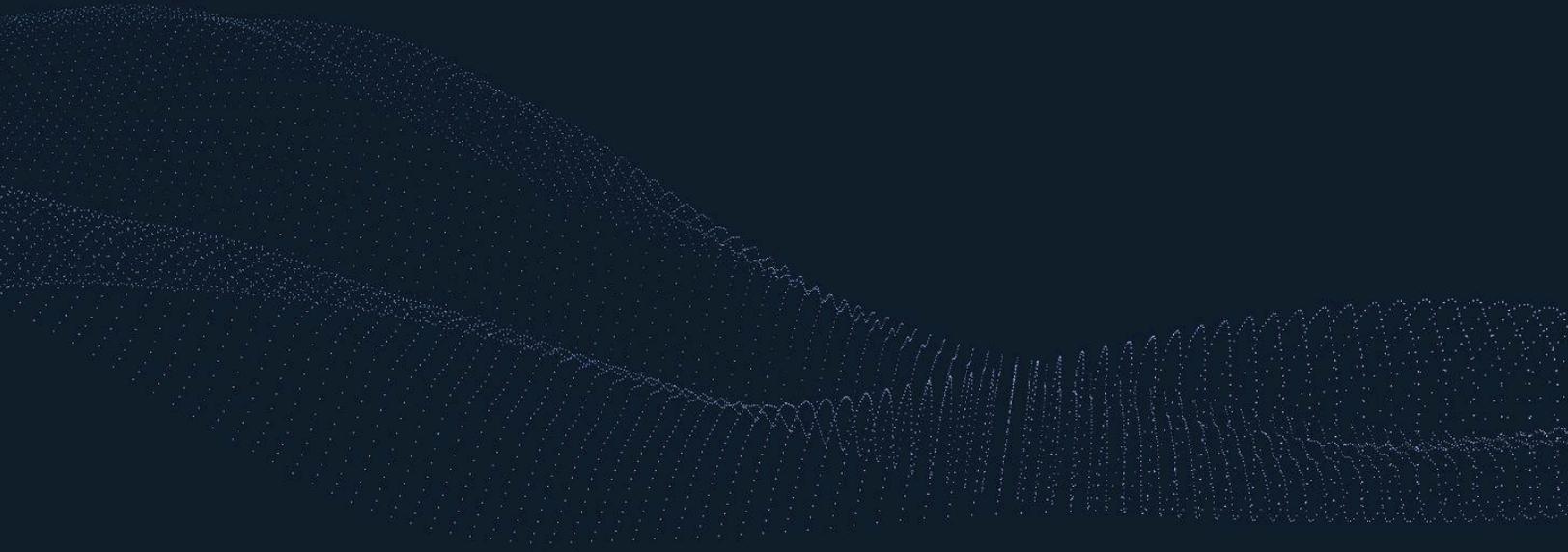